\documentstyle[epsfig,12pt]{article}
\input{psfig.sty}
\newlength{\dinwidth}
\newlength{\dinmargin}
\newlength{\extraspace}
\newlength{\extraspaces}
\newcommand{\be}{\begin{equation}
\addtolength{\abovedisplayskip}{\extraspaces}
\addtolength{\belowdisplayskip}{\extraspaces}
\addtolength{\abovedisplayshortskip}{\extraspace}
\addtolength{\belowdisplayshortskip}{\extraspace}}
\newcommand{\ee}{\end{equation}}
\newcommand{\bdm}{\begin{displaymath}
\addtolength{\abovedisplayskip}{\extraspaces}
\addtolength{\belowdisplayskip}{\extraspaces}
\addtolength{\abovedisplayshortskip}{\extraspace}
\addtolength{\belowdisplayshortskip}{\extraspace}}
\newcommand{\edm}{\end{displaymath}}
\renewcommand{\thefootnote}{\fnsymbol{footnote}}
\def\simlt{\mathrel{\lower2.5pt\vbox{\lineskip=0pt\baselineskip=0pt
           \hbox{$<$}\hbox{$\sim$}}}}
\def\simgt{\mathrel{\lower2.5pt\vbox{\lineskip=0pt\baselineskip=0pt
           \hbox{$>$}\hbox{$\sim$}}}}
\newcommand{\ls}[1]
   {\dimen0=\fontdimen6\the\font
    \lineskip=#1\dimen0
    \advance\lineskip.5\fontdimen5\the\font
    \advance\lineskip-\dimen0
    \lineskiplimit=.9\lineskip
    \baselineskip=\lineskip
    \advance\baselineskip\dimen0
    \normallineskip\lineskip
    \normallineskiplimit\lineskiplimit
    \normalbaselineskip\baselineskip
    \ignorespaces}

\catcode`@=11
\newcount\@tempcntc
\def\@citex[#1]#2{\if@filesw\immediate\write\@auxout{\string\citation{#2}}\fi
  \@tempcnta\z@\@tempcntb\m@ne\def\@citea{}\@cite{\@for\@citeb:=#2\do
    {\@ifundefined
       {b@\@citeb}{\@citeo\@tempcntb\m@ne\@citea\def\@citea{,}{\bf ?}\@warning
       {Citation `\@citeb' on page \thepage \space undefined}}%
    {\setbox\z@\hbox{\global\@tempcntc0\csname b@\@citeb\endcsname\relax}%
     \ifnum\@tempcntc=\z@ \@citeo\@tempcntb\m@ne
       \@citea\def\@citea{,}\hbox{\csname b@\@citeb\endcsname}%
     \else
      \advance\@tempcntb\@ne
      \ifnum\@tempcntb=\@tempcntc
      \else\advance\@tempcntb\m@ne\@citeo
      \@tempcnta\@tempcntc\@tempcntb\@tempcntc\fi\fi}}\@citeo}{#1}}
\def\@citeo{\ifnum\@tempcnta>\@tempcntb\else\@citea\def\@citea{,}%
  \ifnum\@tempcnta=\@tempcntb\the\@tempcnta\else
   {\advance\@tempcnta\@ne\ifnum\@tempcnta=\@tempcntb \else \def\@citea{--}\fi
    \advance\@tempcnta\m@ne\the\@tempcnta\@citea\the\@tempcntb}\fi\fi}
\catcode`@=12

\newcommand{\prd}{{\em Phys.\ Rev.\ }  {\bf D}}

\newcommand{\prl}{{\em Phys.\ Rev.\ Lett.\ }}

\newcommand{\plb}{{\em Phys.\ Lett.\ }{\bf B}}

\newcommand{\zp}{Z. Phys.\ {\bf C}}

\begin{document}
\setcounter{footnote}{1}
\begin{flushright}
BUHEP-00-20\\
SLAC-PUB-8708\\
FERMILAB-Pub-00/299-T\\
\end{flushright}
\vspace{7mm}
\begin{center}
\Large{{\bf 
Semileptonic Form-factors from $B\to K^*\gamma$ Decays
in the Large Energy Limit  
\footnote{Work supported by the Department of Energy 
under contracts DE-AC03-76SF00515, DE-FG02-91ER40676 and 
DE-FG02-95ER40896}}}
\end{center}
\vspace{5mm}
\begin{center}
{
Gustavo Burdman\footnote{burdman@bu.edu}\\
\vspace{0.3cm}
{\normalsize\it Department of Physics, Boston University, }\\ 
{\normalsize
\it 590 Commonwealth Ave., Boston, MA 02215}\\
\vspace{0.4cm}
and \\
\vspace{0.4cm}
Gudrun Hiller\footnote{ghiller@slac.stanford.edu}
}\\
\vspace{0.3cm}
{\normalsize\it Stanford Linear Accelerator Center, }\\ 
{\normalsize
\it Stanford University, Stanford, CA 94309}\\

\end{center}
\vspace{0.50cm}
\thispagestyle{empty}
\begin{abstract}
Making use of the measurement of the $B\to K^*\gamma$ branching ratio 
together 
with the relations following from the limit of high recoil energy,
we obtain stringent constraints
on the values of the form-factors entering in heavy-to-light 
$B\to V\ell\ell'$
processes such as $B\to K^*\ell^+\ell^-$, $B\to K^*\nu \bar{\nu}$ and 
$B\to \rho\ell\nu$ decays.
We show that the symmetry predictions, 
when combined with the experimental information on 
radiative decays, 
specify a severely restricted set of values for the vector and axial-vector
form-factors evaluated at zero momentum transfer, $q^2=0$. 
These constraints can be used to test model calculations 
and to improve our understanding of the $q^2$-dependence of semileptonic
form-factors. We stress that the constraints remain stringent even when 
corrections are taken into account.
\end{abstract}
\newpage

\renewcommand{\thefootnote}{\arabic{footnote}}
\setcounter{footnote}{0}
\setcounter{page}{1}
Semileptonic decays of $B$ mesons play an important role 
in our efforts to put together the pieces of the puzzle 
that the standard model (SM) represents. 
Through decays such as $B\to D^{(*)}\ell\nu$ and $B\to(\pi,\rho)\ell\nu$
some of the fundamental parameters of the SM like the CKM matrix
elements $V_{cb}$ and 
$V_{ub}$ can be measured. 
Further, in modes mediated by flavor changing neutral currents (FCNC)
like e.g.~$B\to K^{(*)}\ell^+\ell^-$ and $B\to K^{(*)}\nu \bar{\nu}$
decays, the short distance structure
of the SM can be tested for contributions from high energy scales, possibly 
due to new physics. 
These exclusive modes have distinct experimental 
signatures in present experiments such as 
$e^+e^-$ $B$ factories (CLEO, BaBar, Belle), as well as the 
$B$-physics programs at high energy colliders
(Tevatron Run II, BTeV and LHC-B).
However, this great potential is somewhat diminished by the fact 
the theoretical predictions for exclusive modes carry 
an uncertainty due to the presence of hadrons in the 
initial and final states. This comes in the form of hadronic matrix elements
parametrized in turn by form-factors, and determined by the non-perturbative, 
long distance dynamics of QCD.

In the last decade a fair amount of progress has been made. Our understanding 
of the behavior of hadrons in the heavy quark limit (HQL) has 
improved since it was discovered that this regime leads to new 
symmetries~\cite{iwsym}. 
Heavy quark symmetries, and the resulting heavy quark effective theory (HQET)
have been of great use in reducing theoretical uncertainty in transitions
where a heavy quark is present in both the initial and the final state hadrons.
This has translated into very small uncertainties in the extraction 
of $V_{cb}$ from $b\to c$ decay modes.
On the other hand, the application of HQET to exclusive heavy-to-light 
transitions has been more limited.

More recently, the large energy limit (LEL), 
which results in additional symmetries impacting 
heavy-to-light decays~\cite{lel1},
has been resuscitated by the authors of Ref.~\cite{lel2}.
In analogy to the HQET, the LEL regime also leads to a controlled 
expansion in the framework of the so-called 
large energy effective theory (LEET). 
In addition to the heavy quark $M\to\infty$ limit,
$E_h\gg\Lambda_{QCD}$ is considered, where $E_h$ denotes the 
final hadronic energy.
It applies to heavy-to-light transitions as the ones we are going to study
in the kinematical range not too close to the zero recoil point.
In the actual $(M,E_h)\to\infty$ limit the matrix elements should
be fully described by perturbative QCD for exclusive processes in the 
Brodsky-Lepage formalism~\cite{bl}. However in practice, $m_b$ is not heavy
enough for the perturbative  approach 
to dominate~\cite{bd91} the form-factors,  
whereas LEET captures the non-perturbative nature of this regime. 
This was shown in Ref.~\cite{lel2} and will be further discussed below.

In addition to those of HQET, LEET enforces new relations among the
relevant form factors.
In this paper we show that combining the well understood heavy
quark spin symmetry (HQSS) with leading order LEL relations and the measured
$B\to K^*\gamma$ branching ratio, 
leads to stringent constraints on the semileptonic form-factors. 
These are particularly important for $B\to V \ell\ell'$ decays, 
with $\ell,\ell'=\ell^{\pm}, \nu$ and $V=K^*,\rho$ 
denoting a light vector meson and enable the determination of the vector and
axial-vector form-factors at zero 
momentum transfer $q^2=0$ in a model independent way. 
We show that corrections in $1/E_h$ and $\alpha_s$ do not affect our results.

We parametrize the hadronic matrix elements over quark bilinears 
relevant for semileptonic
and radiative $B$ meson decays into a vector meson in terms of 
form-factors $V,A_{0,1,2}$ and $T_{1,2,3}$.
These are functions of
$q^2$,
where $q_\mu$ is the momentum transfer into the dilepton
pair and/or the photon in the radiative mode.
In general, the form factors carry also a flavor index depending on
the final quark $q=u,s,(d)$ in the decays under consideration.
They are, however, the same in the SU(3) limit.
We employ the following decomposition
for $B \to V \ell \ell'$ decays of the
``semileptonic'' matrix elements over vector and axial vector currents 
\begin{eqnarray}
\langle V(k,\epsilon)|\bar{q}\gamma_\mu b|B(p)\rangle
&=& \frac{2V(q^2)}{m_B+m_V}\;\epsilon_{\mu\nu\alpha\beta}
\epsilon^{*\nu}p^\alpha k^\beta 
\label{vcurr}\\
\langle V(k,\epsilon)|\bar{q}\gamma_\mu\gamma_5 b|B(p)\rangle
&=& i 2m_V\,A_0(q^2)\frac{\epsilon^*\cdot q}{q^2}q_\mu
+ i (m_B+m_V)\,A_1(q^2)\,\left(\epsilon^*_\mu - \frac{\epsilon^*\cdot q}{q^2}q_\mu
\right) \nonumber\\
& &-i A_2(q^2)\,\frac{\epsilon^*\cdot q}{m_B+m_V}\left( (p+k)_\mu 
-\frac{m_B^2-m_V^2}{q^2}q_\mu\right)~.
\label{vacur}
\end{eqnarray}
and for the FCNC magnetic dipole operator $\sigma_{\mu \nu}$ 
\begin{eqnarray}
\langle V(k,\epsilon)|\bar{q}\sigma_{\mu\nu}(1+\gamma_5) q^\nu b|B(p)\rangle &
=& i\, 2T_1(q^2) 
\epsilon_{\mu\nu\alpha\beta}\epsilon^{*\nu}p^\alpha k^\beta\nonumber\\
& &+ T_2(q^2)\, \left\{ \epsilon_\mu^*(m_B^2-m_{V}^2) - 
(\epsilon^*\cdot p)\,(p+k)_\mu \right\}\nonumber \\
& &+T_3(q^2) (\epsilon^*\cdot p)\,\left\{q_\mu 
-\frac{q^2}{m_B^2-m_{V}^2}\, (p+k)_\mu \right\}~, \label{sig2kst}
\end{eqnarray} 
where $\epsilon_\mu$ denotes the polarization four-vector of the
vector meson $V=\rho,K^*,..$. 
Notice that $T_1(0)=T_2(0)$ and $T_3$ does not contribute to the
amplitude to the radiative decay into an on-shell photon.

\noindent
{\em\underline{The Heavy Quark Limit}:}
In the HQL $m_b\gg \Lambda_{\rm QCD}$ the form factors over the
vector and axial-vector currents are not independent of the dipole
ones.
Instead, they obey the following well known relations 
\cite{iw90,bd91}
\begin{eqnarray}
  T_1(q^2) &=& \frac{m_B^2+q^2-m_V^2}{2m_B}\frac{V(q^2)}
   {m_B+m_V}+\frac{m_B+m_V}{2m_B} A_1(q^2),
\label{rel_2}\\
\hspace*{-0.5cm}\frac{m_B^2-m_V^2}{q^2}\Big[T_1(q^2)-T_2(q^2)\Big] &=&
   \frac{3m_B^2-q^2+m_V^2}{2m_B}\frac{V(q^2)}
   {m_B+m_V}-\frac{m_B+m_V}{2m_B} A_1(q^2),
\label{rel_3}\\
  T_3(q^2) &=& \frac{m_B^2-q^2+3m_V^2}{2m_B} \frac{V(q^2)}{m_B+m_V}+
\frac{m_B^2-m_V^2}{m_B q^2} m_V A_0(q^2) 
\nonumber\\  &\hspace*{-1cm}-&\hspace*{-0.5cm}\frac{m_B^2+q^2-m_V^2}{2m_B q^2}
  \Big[(m_B+m_V)A_1(q^2)-(m_B-m_V)A_2(q^2)\Big].
\label{rel_4} 
\end{eqnarray}
In terms of the symmetries of the HQET, eqns.~(\ref{rel_2}-\ref{rel_4}) 
are a result of the Heavy Quark {\em Spin} Symmetry
that arises in the 
heavy quark limit due to the decoupling of the spin of the heavy 
quark~\cite{iwsym}.

\noindent
{\em\underline{The Large Energy Limit}:}
We now consider the Large Energy Limit (LEL) for heavy-to-light
transitions into a vector meson as the ones we are studying. 
As a result, one recovers the HQSS 
form-factor relations (\ref{rel_2}-\ref{rel_4}), 
but now there are additional new relations among the form-factors defined in 
(\ref{vcurr}-\ref{sig2kst}). 
These will receive corrections that roughly go as 
$(\Lambda_{QCD})/E_h$ and read as \cite{lel2}
\begin{eqnarray}
V(q^2) &=& \left(1+\frac{m_V}{M}\right)\,\xi_{\perp}(M,E)~,\label{le4}\\
A_1(q^2) &=& \frac{2E}{M+m_V} \,\xi_{\perp}(M,E)~,\label{le5}\\
A_2(q^2) &=&\left(1+\frac{m_V}{M}\right)\,\left\{
\xi_\perp(M,E) - \frac{m_V}{E}\xi_\parallel(M,E)\right\}~,
\label{le6}\\
A_0 (q^2)&=& \left(1-\frac{m_V^2}{ME}\right)
\xi_\parallel(M,E) + \frac{m_V}{M}\xi_\perp(M,E)~,
\label{le7}
\end{eqnarray}
and
\begin{eqnarray}
T_1(q^2) &=& \xi_\perp(M,E)~,\label{t1}\\
T_2(q^2) &=& \left(1-\frac{q^2}{M^2-m_V^2}\right) \xi_\perp(M,E)~,
\label{t2}~,\\
T_3(q^2)&=&\xi_\perp(M,E) - \frac{m_V}{E}\left(1-\frac{m_V^2}{M^2}\right)\xi_\parallel(M,E)~.
\label{t3}
\end{eqnarray}
It is apparent from eqns.~(\ref{le4})-(\ref{t3}) that, in the LEL regime, 
the $B\to V\ell\ell'$ decays are described by only two form-factors:
$\xi_\perp$ and $\xi_\parallel$, instead of the seven apriori independent
functions in the general Lorentz invariant ansatz of the matrix
elements. Here, 
$\xi_\perp$ and $\xi_\parallel$ are functions of the heavy mass
$M$ and the hadronic energy $E$, and refer to the transverse and longitudinal
polarizations, respectively.

This simplification leads to new relations among the form-factors. 
For instance, the ratio of the vector form-factor $V$
to the axial-vector form-factor $A_1$, 
\begin{equation}
R_V(q^2)\equiv\frac{V(q^2)}{A_1(q^2)}=\frac{(m_B+m_V)^2}{2E_Vm_B}~,
\label{rv}
\end{equation}
is independent of any of these unknown, non-perturbative functions
$\xi_{\perp,\parallel}$
and is determined by
purely kinematical factors. Here, $E_V=(m_B^2+m_V^2-q^2)/(2 m_B)$ denotes
the energy of the final light
vector meson. A similar relation holds 
for $T_1$ and $T_2$, since they both are also proportional to the  
``transverse'' form-factor $\xi_\perp$. 
As we will see below, these predictions have important consequences 
for observables at large recoil energies (low $q^2$). 

The leading corrections to expressions such as eqn.~(\ref{rv}) are 
expected to be of order ${\cal O}(\Lambda_{\rm QCD}/2E_V)$, so if
$q^2\to 0$, then $E_V \simeq  m_B/2$ and 
the corrections are expected to be typically below $10\%$. 
Additional corrections from perturbative QCD arise through the 
exchange of hard gluons~\cite{beneke}, which are also small and below 
the $10\%$ level. 
This confirms the result from Ref.~\cite{bd91} derived for 
$B\to\pi\ell\nu$: although pQCD should {\em formally} dominate
the $M\to\infty$ limit, this is not what actually 
happens for $M\simeq m_b$, namely
hard gluon corrections to the LEL relations are small.

Furthermore, the ratio $R_V$ defined in eqn.~(\ref{rv}), 
does not receive $\alpha_s$ corrections. The reason for this, as well as
the physical picture behind expression~(\ref{rv}), 
becomes clear once we look at
the transverse helicity amplitudes for a generic $B\to V\ell\ell'$ transition. 
Making use of the HQSS relations (\ref{rel_2})~and (\ref{rel_3}), 
these can be written as
\begin{equation}
 H_\pm = {\cal F} \;(V\mp\frac{(m_B+m_V)^2}{2m_Bk_V}A_1)~,
\label{hel}
\end{equation}
where $\cal F$ is a factor depending on the mode under consideration
(e.g.~Wilson coefficients, coupling constants, etc...)
and $k_V$ is the momentum of the vector meson. 
Thus, we see from the form of $R_V$ in the large energy limit, 
that the ``$+$'' helicity vanishes $H_+=0$ in the LEL regime, up to residual 
terms of order $m_V^2/2E_V^2$. This is not a surprise: in the 
limit of an infinitely heavy quark decaying into a light quark, 
the helicity of the latter is ``inherited'' by the final vector meson.
In the SM, the $(V-A)$ structure in semileptonic decays is reflected
in the dominance of the $H_-$ transverse helicity. 
On the other hand, the amplitude to flip the helicity of the fast 
outgoing light quark is suppressed by $1/E_h$. 
This is also the reason why $\alpha_s$ corrections from hard
gluon exchange between the spectator quark and the fast light quark
do not affect eqn.~(\ref{rv}): they are not helicity-changing.  
By the same reasoning, the same is true for the ratio
of $T_1$ and $T_2$.

Finally, we point out that the expression (\ref{rv}) for $R_V$
is expected to hold in most relativistic quark models that compute
the form-factors at $q^2=0$. This is the case 
because these model calculations, 
although rather uncertain in the absolute value of each form-factor 
{\em per se}, are likely to respect the helicity conservation
property of the fast outgoing light quark. The overall uncertainty
in each form-factor comes in as the overlap of meson 
wave-functions, and largely vanishes in the ratio $R_V$.
This was found in Ref.~\cite{gbafb} in the context of predictions
for the forward-backward asymmetry in $B\to K^*\ell^+\ell^-$, 
where it was shown that the position of the zero of the asymmetry only depends
on $R_V$. Since the zero is located in the low $q^2$ region, 
(around $q^2=3 \mbox{GeV}^2$ in the SM)
and in the region of validity of the LEL, one can use eqn.~(\ref{rv})
to predict $R_V$ and the position of the asymmetry zero  
with very small hadronic uncertainties \cite{lcsr}.

\noindent
{\em\underline{Constraints on Semileptonic Form-factors at $q^2=0$}:}
We now extract the magnitude of the form factor $T_1(0)$ from the
branching ratio of $B\to K^*\gamma$ decays. It is customary to normalize the 
exclusive to the inclusive $B\to X_s\gamma$ branching ratio, thus
eliminating the uncertainties from the CKM factor $V_{tb}V_{ts}^*$
and the SM short distance Wilson coefficient.
This results in the ratio\footnote{
In fact, even in the presence of physics beyond the SM, the Wilson 
coefficient cancels in the ratio $R_\gamma$ as long as there is no 
sizeable contribution to the ``flipped chirality'' dipole
operator $\bar s_R\sigma_{\mu\nu}b_L $. 
} 
\begin{equation}
R_\gamma\equiv \frac{Br(B\to K^*\gamma)}{Br(B\to X_s\gamma)}
=\frac{m_B^3}{m_b^3} (1- (\frac{m_{K^*}}{m_B})^2 )^3 |T_1(0)|^2 ~,
\label{rgamma}
\end{equation}
which can be evaluated using data~\cite{bkexp,cleobsg}:
\begin{eqnarray}
Br(B\to K^{*}\gamma) &=& (4.25\pm0.55\pm0.29)\times10^{-5}\nonumber\\
Br(B\to X_s\gamma) &=& (3.15\pm 0.35\pm0.32\pm0.26)\times10^{-4}~.
\nonumber
\end{eqnarray}
Note that the exclusive branching ratio reflects a weighted average of the 
charged and neutral modes~\cite{thanks}.
We obtain $R_\gamma=0.135\pm0.030$,
leading to $|T_1(0)|=0.333\pm 0.043$.
Here we employed an on-shell $b$-quark mass in the conservative range of
$m_b=(4.8\pm 0.2)$~GeV to evaluate the phase space
factor $m_b^3$ from the inclusive decay in eqn.~(\ref{rgamma}).
However, the dominant uncertainty in the extraction of $T_1(0)$ stems
from the experimental input in $R_\gamma$. 

\begin{figure}[h]
\leavevmode
\centering
\epsfig{file=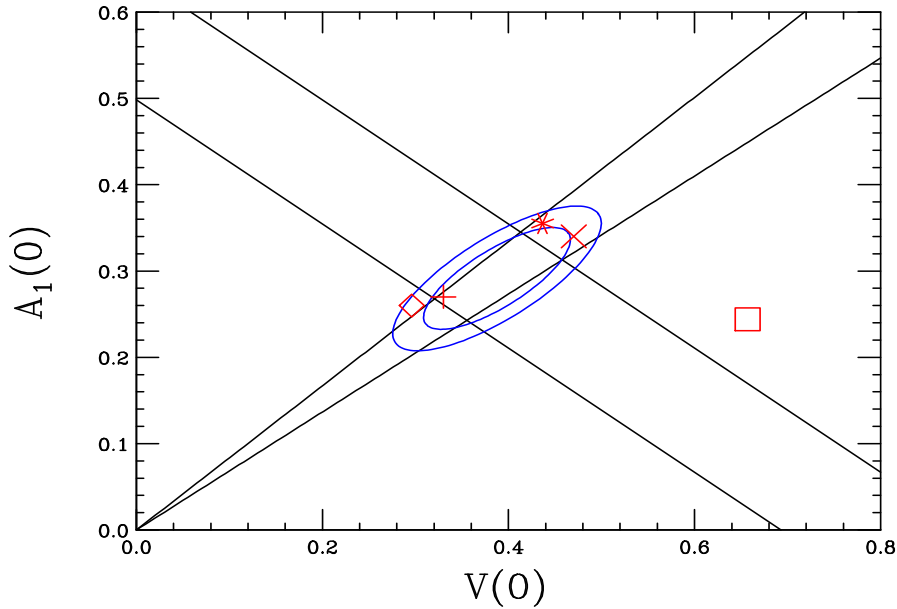,width=8cm,height=11.8cm,angle=90}
\caption{Constraints on the semileptonic form-factors $V(0)$ and $A_1(0)$ from 
$B\to K^*\gamma$ data plus HQSS (thicker band) together with the
relation from the LEL (thiner band).
The ellipses correspond to $68\%$ and $90\%$ confidence level intervals. 
Central values of model predictions are also shown and correspond to 
BSW~\cite{bsw} (vertical cross), ISGW2~\cite{isgw2} (diamond),
 MS~\cite{ms} (star), LCSR~\cite{lcsr} (diagonal cross) and
LW~\cite{lw} (square), respectively.}
\label{vvsa1}
\end{figure}

We recall that HQSS relates form-factors of matrix elements of magnetic
dipole operators to those of semileptonic currents. 
At $q^2=0$ eqn.~(\ref{rel_2}) can be written as
\begin{equation}
A_1(0)=\frac{2 m_B}{m_B+m_{K^*}} \, T_1(0)-\frac{m_B-m_{K^*}}{m_B+m_{K^*}}
\,V(0)~,
\label{expconst}
\end{equation} 
Using for $T_1(0)$ in (\ref{expconst}) 
the value extracted from the measurement of $R_\gamma$   
translates into a constraint 
in the $[V(0),A_1(0)]$ plane,
which is displayed in Fig.~\ref{vvsa1} (thicker band). 
On the other hand  the ratio of these form-factors, $R_V$,  
which in the large energy limit is given by eqn.~(\ref{rv}) ,  
constitutes another constraint.
The LEL constraint (cone in Fig.~\ref{vvsa1}) is plotted 
assuming a $10\%$ error in the ratio, 
which we believe to be conservative. 
In fact, 
the typical size of this error is 
${\cal O}(\Lambda_{\rm QCD}/2E_h^{\rm max.})\simeq 6\%$. By nearly doubling
its size we expect to safely account for the fact that this
is a non-gaussian error. 
The intersection of the HQSS plus $B\to K^* \gamma$ data constraint
with the LEL expression for $R_V$ 
leads to the two ellipses corresponding to the $68\%$ (solid) and
$90\%$ (dashed) confidence intervals (i.e. $1.5\sigma$ and $2.1\sigma$ 
respectively)
\footnote{To obtain the ellipses from linear fitting, we approximate 
the cone by a band with thickness given by that of the 
cone at the intersection with the HQSS constraint.}. 
Our fit results in 
\begin{equation}
V(0)=0.39\pm0.06~,~~~~~~~~~~~~~~~~~~~A_1(0)=0.29\pm0.02~. 
\label{result}
\end{equation}

We compare our findings for $V(0)$ and $A_1(0)$ with
several model predictions in 
Fig.~\ref{vvsa1}. For illustration, we take 
the Bauer-Stech-Wirbel (BSW) model from Ref.~\cite{bsw}
(cross), the modified 
version of the Isgur-Scora-Grinstein-Wise (ISGW2) model from 
Ref.~\cite{isgw2} (diamond), 
a recent relativistic constituent
quark model prediction by  Melikhov and Stech (MS) \cite{ms} (star),
the recent calculation in the Light Cone QCD Sum Rule (LCSR) formalism
of Ref.~\cite{lcsr} (diagonal cross)  and 
the prediction by Ligeti and Wise (LW) from Ref.\cite{lw} (square).
We see that relativistic constituent quark models, which directly 
compute the form-factors at $q^2=0$, fall close to the constraint. 
This is the case with the models of Refs.\cite{bsw,lcsr,ms}. 
The ISGW2 prediction, although slightly outside the $68\%$ C.L. contour, 
fares rather well, probably not in small measure 
due to the relativistic corrections
added with respect to the non-relativistic ISGW model~\cite{isgw}. 

On the other hand, the LW prediction (square in Fig.~\ref{vvsa1}) 
appears to be excluded. It is based on
$D\to K^*\ell\nu$ data, heavy quark flavor symmetry and
assuming monopole $q^2$-dependence of the form-factors. 
This latter assumption is needed in order to 
extrapolate from the small recoil energies 
of charm decays ($E_h\leq1.14~$GeV) to the $q^2=0$
region in $B$ decays, which corresponds to $E_{K^*}\simeq m_B/2~$.
Although the heavy quark flavor symmetry is expected to be affected
by large corrections, these are unlikely to produce such a 
shift with respect to the symmetry predictions.
The assumption of monopole behavior for the $q^2$-dependence  on the 
other hand, is not well justified far away from the zero recoil point. 
In fact, it is known that in the deep Euclidean region
form-factors should match to the pQCD predictions.
For vector form-factors this asymptotic behavior for  
$q^2\ll 0$ (but still $|q^2|<m_B^2\,\ln(m_B^2/\Lambda_{\rm QCD})$) 
is that of a dipole~\cite{bd91,monsup}.
Thus, it is possible that around $q^2=0$ the $q^2$-dependence is suppressed
with respect to that of a monopole, even if it is not 
completely a dipole~\cite{monsup}. On the other hand, 
$A_1(q^2)$ may not be as suppressed due to the  additional 
factor~\cite{a1q2} $(1-q^2/m_B^2)$, which is also present in $T_2(q^2)$. 
The suppression of $V(q^2)$ could bring the 
LW prediction into line with the constraint of Fig.~\ref{vvsa1}.

We point out that the LEL relations alone are sufficient to determine
$[V(0),A_1(0)]$ from the $b\to s \gamma$ data and (\ref{rgamma}),
respectively, without employing the HQSS relation~(\ref{expconst}).
Feeding eqn.~(\ref{t1}) into  eqns.~(\ref{le4}) and (\ref{le5})
yields 
$V(0)=0.39\pm 0.05$ and $A_1(0)=0.29 \pm 0.04$, 
in agreement with our previous result eqn.~(\ref{result}).
Further, LEET predicts a simple relation between $T_3$ and $A_2$,
namely $T_3(0)/A_2(0)=(m_B-m_V)/m_B + {\cal{O}}(m_V^2/m_B^2) \sim 0.83$
\footnote{For comparison, using the central values from Ref.~\cite{lcsr} we
obtain for this ratio the value $0.92$, whereas following the procedure of 
Ref.\cite{lw} the obtained value is $1.36$.}.

With the use of $SU(3)$ flavor symmetry, the constraints obtained
above can be directly 
imposed on the analogous form-factors entering in $B\to\rho\ell\nu$ 
decays. Corrections to the $SU(3)$ limit at large values of the 
hadronic recoil energies are expected to be of the order of~\cite{gbafb}
\begin{equation}
\delta\simeq \frac{(m_s-m_q)}{E_h}~, 
\label{su3}
\end{equation}
with $m_q$ the $u,d$ constituent quark mass.
Thus, for typical values of the constituent quark masses, 
we expect the $SU(3)$ corrections relevant to the 
constraints in Fig.~\ref{vvsa1} to be below $10\%$.

\noindent
{\em\underline{Conclusions}:}
We have derived stringent constraints on the 
vector and axial-vector form-factors $V(q^2)$ and $A_1(q^2)$  
entering in $B\to K^* \ell^+\ell^-$, $B\to K^*\nu\bar\nu$ and 
(in the $SU(3)$ limit) $B\to\rho\ell\nu$ decays. 
These apply to 
the highest recoil energy of the vector meson, i.e. $q^2=0$.
We emphasize that these constraints, which are summarized in
Fig.~\ref{vvsa1}, come exclusively from: 
\begin{itemize}
\item Data on $B\to K^*\gamma$ and 
$B\to X_s\gamma$ branching ratios, 

\item Heavy Quark {\em Spin} symmetry 
(in this case (\ref{rel_2})) and

\item the Large Energy Limit (in particular eqn.~(\ref{rv}), the ratio 
of (\ref{le4}) and (\ref{le5})).
\end{itemize}
Of these three ingredients, the first one is derived 
from experimental measurements,
and the second one is a well established symmetry relation with corrections
well below the experimental errors in the branching ratios. 
As discussed above, the third element is a direct consequence of 
the helicity conservation property of the strong interactions,
which implies that in the LEL, helicity flipping is down by 
$(\Lambda_{\rm QCD}/2E_h$). This leads to a purely kinematical 
expression for the ratio of the vector-to-axial-vector form
factor $R_V$, eqn.~(\ref{rv}), 
valid to leading order in the $1/E_h$ expansion.
We thus conclude that these constraints are fairly solid and model independent.
In any event, a rigorous treatment of the leading corrections
in LEET is still lacking and should be undertaken.
On the other hand, the experimental errors in the measurements of 
both the exclusive and inclusive radiative decays could be substantially
reduced in the $B$ factory era, leading to an even more stringent 
constraint in Fig.~\ref{vvsa1}.

Lattice gauge theory calculations of the form-factors have made
great progress in recent years~\cite{lattice}.
However, they are confined to the region of low recoil energy. 
The constraints derived here allow an extrapolation from 
this region down to low values of $q^2$, without {\em ad hoc } assumptions 
about the $q^2$-dependence of the form-factors.

In the LEL, $SU(3)$ corrections are at most of order $10 \%$, allowing 
our constraints to be also imposed on the $B \to \rho\ell\nu$ form-factors.
The use of LEET also results in similar results in 
$B\to P$ transitions (with $P=\pi,K,{\rm etc.}$), as well as in 
baryon decays such as $\Lambda_b\to\Lambda\gamma$ and 
$\Lambda_b\to\Lambda\ell^+\ell^-$, where only one form-factor
is needed to determine the hadronic matrix elements. 

The precise knowledge of the form-factors at $q^2=0$ gives us a
handle to understand their $q^2$-dependence, as well as testing model 
calculations.
The reduction of the theoretical uncertainties inherent to the 
description of exclusive semileptonic heavy-to-light decays 
such as $B\to\rho\ell\nu$ 
facilitates the clean extraction of the SM parameter $V_{ub}$.  
At the same time, it allows us to test the short distance structure
of the SM for new physics contributions to FCNC mediated decays
such as $B\to K^*\ell^+\ell^-$ and $B\to K^*\nu\bar\nu$~\cite{g3}. 

\noindent
{\em\underline{Acknowledgments}:} 
We thank
Frank Wuerthwein and David Jaffe for correspondence on 
$B \to K^* \gamma$ data from CLEO and Stan Brodsky,  Lance Dixon,
Andre Hoang and
Michael Peskin for useful discussions
and computing support.
We also thank Zoltan Ligeti
and Mark Wise for discussions and comments on the manuscript. 
G.B. thanks the LBL theory group and 
G.H. thanks the Fermilab theory group 
for their hospitality during the completion of 
this work. 



\begin{thebibliography}{99}
\bibitem{a1q2} This is necessary if the scaling of the form-factor with 
the heavy mass is to be matched at both zero as well as maximum  recoil.
See for instance P. Ball and V. M. Braun,  
\prd55, 5561 (1997).
\end{thebibliography}
\end{document}